\documentclass[aip,pof,reprint,onecolumn]{revtex4-1}
\usepackage{amsmath}
\usepackage{graphicx}
\usepackage[dvipsnames]{color}
\usepackage{hyperref}
\usepackage{psfrag}
\usepackage{stmaryrd}
\usepackage{amssymb}
\usepackage{wasysym}
\usepackage[latin1]{inputenc}
\usepackage{bm}
\makeatletter
\@ifpackageloaded{textcomp}
  {\newcommand{\degre}{\textdegree}}  %% new-style LaTeX degree
  {\newcommand{\degre}{$^\circ$}}     %% old-style TeX degree
\makeatother

\usepackage{geometry}
\usepackage[normalem]{ulem}

\DeclareSymbolFont{symbolsC}{U}{txsyc}{m} {n}
\SetSymbolFont{symbolsC}{bold}{U}{txsyc}{bx}{n}
\DeclareMathSymbol{\medcirc}{\mathbin}{symbolsC}{7}

\begin{document}

\title{Influence of the multipole order of the source on the decay of an inertial wave beam in a rotating fluid}

\author{Nathana\"el Machicoane}
\affiliation{Laboratoire FAST, CNRS, Universit\'e Paris-Sud,
Orsay, France}
\author{Pierre-Philippe Cortet}
\affiliation{Laboratoire FAST, CNRS, Universit\'e Paris-Sud,
Orsay, France}
\author{Bruno Voisin}
\affiliation{Laboratoire LEGI, CNRS, Universit\'e Grenoble Alpes, Grenoble, France}
\author{Fr\'{e}d\'{e}ric Moisy}
\affiliation{Laboratoire FAST, CNRS, Universit\'e Paris-Sud,
Orsay, France}

\date{\today}

\begin{abstract}

We analyze theoretically and experimentally the far-field viscous
decay of a two-dimensional inertial wave beam emitted by a harmonic line source
in a rotating fluid. By identifying the relevant conserved quantities
along the wave beam, we show how the beam structure and decay exponent
are governed by the multipole order of the source.
Two wavemakers are considered experimentally,
a pulsating and an oscillating cylinder, aiming to produce a
monopole and a dipole source, respectively. The relevant conserved quantity which
discriminates between these two sources is the instantaneous
flowrate along the wave beam, which is non-zero for the monopole
and zero for the dipole. For each source the beam structure and decay exponent,
measured using particle image velocimetry, are in good agreement with the predictions.

\end{abstract}

\maketitle

\section{Introduction}

In rotating fluids, the restoring action of the Coriolis force
allows for the propagation of anisotropic, transverse, circularly
polarized waves called inertial waves.\cite{GreenspanBook} These
waves are of fundamental interest for geo- and astrophysical
flows:\cite{Aldridge_Toomre_1969,Aldridge_Lumb_1987,Suess_1971}
they can for instance be excited in the fluid core of planets by
tidal motions, precession or
libration.\cite{Kerswell1995,Rieutord2001,Kida2011} Internal gravity
waves in stratified fluids, relevant to the ocean and the
atmosphere, share a number of properties with inertial
waves.\cite{Lighthill1978,Pedlosky1987} Internal waves, or mixed
internal--inertial waves when rotation and stratification effects
are of comparable magnitude,\cite{Peat1978,Peacock2005} can also
be excited in the ocean by the interaction of tides with
topography.\cite{Wunsch2004,Garrett2007,Echeverrri2010,Swart2010}

\begin{figure}[t]
\centerline{\includegraphics[width=0.9\columnwidth]{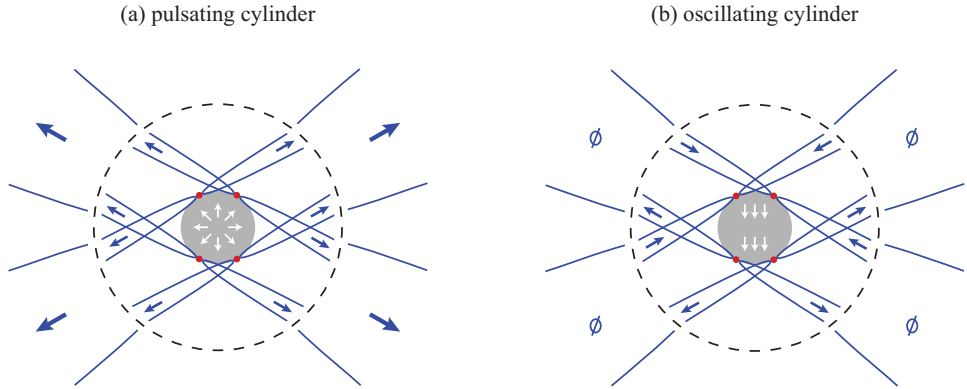}}
\caption{(Color online) Sketch of the far-field wave beams
resulting from a combination of parent beams emitted from the
critical lines, for (a) a
pulsating cylinder and (b) an oscillating cylinder.
The size of the disturbance is assumed much larger than the thickness
of the Ekman boundary layer.
The dashed circles represent the separation
between bimodal and unimodal regions. Eight wave
beams are emitted from four critical lines (red points), which
behave as local sources of non-zero instantaneous flowrate. Far
from the cylinder, two parent
beams propagating in the same direction combine and
form a unique beam. The instantaneous flowrate of the
resulting merged beam is non-zero when the two parent beams
are in phase (case a), and is zero when they are out of phase
(case b).}\label{fig:sketch}
\end{figure}

We focus in this paper on the scaling of the viscous decay of a wave beam emitted
by a line source in a uniformly rotating fluid. Although this
problem has received much attention, the influence of the
multipole order of the source has not been addressed so far.
Interestingly, whereas the growth of the wave beam thickness,\cite{Walton1975}
$\delta(x) \sim x^{1/3}$ (with $x$ the distance from the source),
is independent of the multipole order of the source,
the decay of the wave amplitude depends on which quantity is conserved along the wave
beam (flowrate, momentum, or higher order moment), which 
is directly governed by the multipole order of the source.
Such dependence
was discussed in the case of internal waves produced by a point
source in stratified fluids by Voisin.\cite{Voisin2003} Here
we address this problem for a two-dimensional inertial wave beam
produced by a line source, extending the simple far-field
quasi-parallel approach of Cortet \textit{et al.}\cite{Cortet2010}
to a source of arbitrary multipole order. We show that the wave
decay is steeper as the multipole order of the source increases,
which implies that the far-field decay of a wave beam emitted by an
arbitrary source is dominated by its lowest multipole component.

Inertial wave beams emitted from localized sources are relevant to
a broad range of laboratory and natural flows, including local forcing by
a wavemaker immersed in the fluid, but also global forcing such as
precession, libration or tidal motion acting at the scale of the fluid
domain. This is because in all cases wave beams are emitted from critical
lines, where the local slope of the solid boundaries equals the
propagation angle of the wave.  Along such critical lines the
oscillating boundary layer erupts and forms oscillating
beams in the bulk of the flow.\cite{GreenspanBook,Kerswell1995,Hollerbach1995,Rieutord2001,
Rieutord2002,Calkins2010}
In confined fluid domains such beams reflect and, in the presence
of sloping boundaries, may focus on wave
attractors.\cite{Tilgner1999,Maas2001,Rieutord2001,Manders2003} 
The eruption at critical lines produces two types of wave beams, associated with different magnitudes and scaling laws, propagating in planes tangent and non-tangent to the solid boundary.\cite{Kerswell1995} Wave emission in the tangent plane, developing only along convex boundaries, is stronger:
this is the case for conical inertial wave beams emitted from
the inner core of a rotating spherical shell,\cite{Kerswell1995,Hollerbach1995}
or for internal
wave beams excited by oceanic internal tides on the edge of continental
shelves or ridges.\cite{Wunsch2004,Garrett2007,Gostiaux2007,Echeverrri2010}
Wave emission in the non-tangential plane, both at critical
lines of convex or concave slope (for instance on
the outer sphere of a rotating spherical shell\cite{Kerswell1995,Hollerbach1995,Rieutord2001} or at the sloping bottom
of a ridge\cite{Swart2010}), is of weaker amplitude.
Emission of inertial wave beams is also observed from horizontal edges in containers such as a cylinder\cite{McEwan1970,Duguet2006} or a parallelepiped.\cite{Boisson2012b}

A key property of an inertial wave beam spawned from an erupting boundary layer
is its non-zero instantaneous flowrate: it can be modeled in the far field
as originating from a monopole
line source. Depending on the topology of the fluid domain,
in particular on the distribution and 
relative phases between such elementary monopole sources,
different wave beams propagating in the same direction
may combine and form far-field beams of either
non-zero or zero instantaneous flowrate, therefore corresponding
to an effective monopole or a higher order multipole source.
Note that such combination of beams requires propagation over a
distance much larger than the separation
between the sources, a requirement which is usually not satisfied in geo-
and astrophysical situations.

We restrict in the following to wave beams produced by effective line sources
surrounded by the fluid.
In practical situations, such line source corresponds
to a two-dimensional convex disturbance,
say a cylinder, defining four critical lines  (Fig. 1).
If the cylinder is pulsating (Fig.
1a), the periodic emission and suction of mass from these critical lines
is in phase, so the far-field merged beams have
non-zero flowrate: this defines an effective monopole source.
On the other hand, if the cylinder is
oscillating  (Fig. 1b), the parent beams
propagating in the same direction are out of phase, resulting in
far-field merged beams of zero flowrate: this defines an effective dipole source.
The pulsating and
oscillating cylinders are therefore generic configurations to
investigate the influence of the multipole order of a line source on
the properties of the far-field wave beams.

Most of the experiments investigating wave beams emitted by a local
disturbance, both in
stratified\cite{Mowbray1967,Thomas1972,Flynn2003,Zhang2007,Gostiaux2007,Voisin2011}
and rotating\cite{Messio2008,Cortet2010} fluids, are based on
oscillating wavemakers (with the exception of
Makarov {\it et al.}\cite{Makarov1990} who report results from a
pulsating cylinder in a stratified fluid). The resulting far-field
wave beams are therefore distinct from  those 
produced from an erupting boundary layer in a globally forced fluid domain.
The aim of the present paper
is to compare the far-field properties of inertial wave beams
emitted from a pulsating and an oscillating cylinder in a rotating tank,
aiming to produce a monopole and a dipole source.
Velocity measurements in the wave beam are achieved using
two-camera multi-resolution particle image velocimetry,
ensuring a good resolution both in the near and far fields.
The two sources produce distinct decay exponents and wave beam profiles,
in good agreement with the theoretical predictions.

\section{Theoretical background}\label{sec:theory}

\subsection{Dispersion relation and viscous spreading}

The geometrical properties of inertial waves follow from their
dispersion relation,\cite{GreenspanBook}
\begin{equation}
\sigma = \pm \frac{2 \mathbf{\Omega} \cdot \mathbf{k}}{|\mathbf{k}|} = 2 \Omega \cos \theta ,
\label{eq:dr}
\end{equation}
with $\sigma>0$ the wave frequency and $\theta$ the polar angle
between the wave vector $\mathbf{k}$ and the vertical rotation
vector $\mathbf{\Omega}$. Waves emitted from a localized harmonic
disturbance propagate energy in directions making an angle $\pm
\theta$ to the horizontal, along two cones for a point source and
along four plane beams for a line source normal to
$\mathbf{\Omega}$. We restrict in the following to the line source
configuration, for which the spatial decay of the wave is purely
governed by viscosity. In each wave beam, fluid particles describe
anticyclonic circular translations in the tilted plane normal to
$\mathbf{k}$. Since only the orientation of $\mathbf{k}$ is
prescribed by the dispersion relation (\ref{eq:dr}) but not its
magnitude, the characteristic sizes of the wave (wave length, beam
thickness) are governed by the boundary conditions and viscosity.

The viscous spreading of the wave beam results from the combination of the
energy propagation in the longitudinal direction $x$ and its
diffusion in the lateral direction $z$ (see Fig.~\ref{fig:contour}).
Its scaling can be obtained from a classical boundary-layer argument:
During a time $t$, the wave energy spreads laterally over a distance
$\delta \simeq \sqrt{\nu t}$, with $\nu$ is the kinematic viscosity,
and propagates over a distance $x = c_g t$, where $c_g$ is the group
velocity. Evaluating $c_g = (\sigma/k) \tan \theta$ for the dominant
wave number at a distance $x$ from the source, $k \sim \delta^{-1}$,
simply yields
\begin{equation}
\label{eq:del}
\delta(x) \sim \ell^{2/3} x^{1/3},
\end{equation}
where we introduce the viscous scale
\begin{equation}
\label{eq:ell}
\ell = (\nu / \sigma \tan \theta)^{1/2}.
\end{equation}
Although the scaling of the wave amplitude strongly
depends on the multipole order of the source,
the scaling of the beam thickness (\ref{eq:del}) is
independent of the nature of the source, provided that $\delta(x)$
is much larger than the size of the source.

\subsection{Boundary layer equations}

We derive now the similarity solutions for a viscous
2D inertial wave beam emitted by a harmonic line source, focusing
on the spatial decay of the wave amplitude and its dependence on
the multipole order of the source. The derivation follows that of
Thomas and Stevenson\cite{Thomas1972} for internal waves in
stratified fluids. We use the velocity--vorticity formulation of
Cortet {\it et al.},\cite{Cortet2010} generalized here to a source
of arbitrary order.

\begin{figure}
\centerline{\includegraphics[scale=.6]{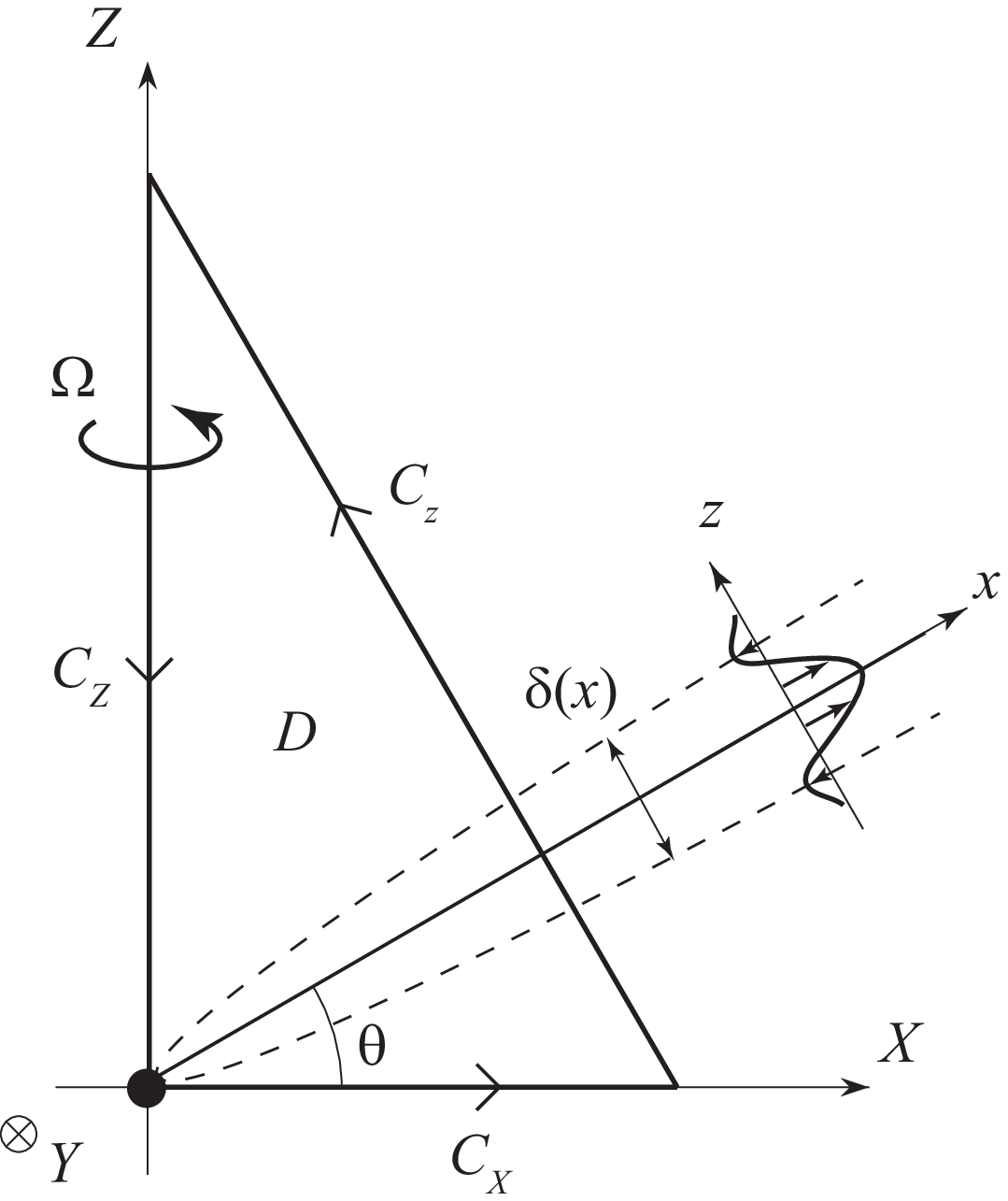}}
\caption{Two-dimensional wave beam emitted from a line source at
$(X=0,Z=0)$ in a fluid rotating about the $Z$ axis. Only the beam
propagating in the first quadrant ($X>0$, $Z>0$) is shown.
The contour $C$ shows the frontiers of the domain of integration
$D$ used in Sec.~\ref{sec:MOandMS}.} \label{fig:contour}
\end{figure}

We consider a line source along the $Y$ axis, of angular frequency
$\sigma$, in a fluid rotating at rate $\Omega$ about the $Z$ axis
(Fig.~\ref{fig:contour}). The four wave  beams emitted by the
source being invariant along $Y$,  energy propagates in the
$(X,Z)$ plane. In the following, we consider only the wave beam
propagating in one given quadrant. We start from the linearized
vorticity equation in the rotating frame
\begin{equation}
\label{eq:dtw}
\partial_t \boldsymbol{\omega} = (2 \mathbf{\Omega} \cdot \nabla) \mathbf{u} + \nu \nabla^2 \boldsymbol{\omega},
\end{equation}
with $\mathbf{u}$ the velocity and $\boldsymbol{\omega} = \nabla
\times \mathbf{u}$ the vorticity. We project (\ref{eq:dtw}) on the
local frame of the far-field wave beam, $(\mathbf{e}_x,
\mathbf{e}_y, \mathbf{e}_z)$, with $\mathbf{e}_x$ aligned with the
group velocity, making angle $\theta = \cos^{-1} (\sigma /
2\Omega)$ with the horizontal, and assume that the flow inside the
wave beam is quasi-parallel (boundary layer approximation), i.e.,
such that $|u_x|, |u_y| \gg |u_z|$; $|\omega_x|, |\omega_y| \gg
|\omega_z|$ and $\nabla^2 \simeq \partial_z^2$. We introduce the
complex velocity and vorticity fields in the tilted plane $(x,y)$,
$U = u_x + i u_y$ and $W = \omega_x + i \omega_y \simeq i
\partial_z U$. Searching for harmonic solutions of the form $U =
U_0 e^{-i \sigma t}$, Eq.~(\ref{eq:dtw}) writes
\begin{equation}
\partial_x U_0 + i \ell^2 \partial_z^3 U_0 = 0.
\label{eq:b}
\end{equation}
This equation admits similarity solutions as a function of the
reduced transverse coordinate $\eta = z/(x^{1/3} \ell^{2/3})$ of
the form
\begin{equation}
U_0 (x,z) = \tilde U_0 \left( \frac{\ell}{x} \right)^a
f(\eta), \label{eq:u0a}
\end{equation}
where $\tilde U_0$ is a complex velocity scale and $a > 0$ the
decay exponent to be determined. The solution considered in Cortet
{\it et al.}\cite{Cortet2010} was derived for the particular case
$a=1/3$.\cite{footnote} Inserting Eq.~(\ref{eq:u0a}) in Eq.~(\ref{eq:b}) yields
\begin{equation}
3 f''' + i \eta f' + 3 i a f = 0.
\label{eq:ts}
\end{equation}
Solutions of this equation, first given by Moore and
Saffman\cite{Moore1969} for the problem of a vertical steady shear
layer ($\theta=\pi/2$), and later by Thomas and
Stevenson\cite{Thomas1972} in the context of internal waves, are
\begin{equation}
f_m (\eta) =c_m(\eta) + is_m(\eta) = \int_0^\infty K^m e^{-K^3} e^{iK\eta} dK,
\label{eq:fm}
\end{equation}
with $c_m$ and $s_m$ real functions, even and odd respectively.
The properties of these functions were considered in detail by
Voisin.\cite{Voisin2003} Integrating Eq.~(\ref{eq:fm}) by parts
gives $3 f_{m+3} - i\eta f_{m+1} - (m+1) f_m = 0$ for $m>-1$,
which (using $f'_m = i f_{m+1}$) yields
\begin{equation}
3 f'''_m + i\eta f'_m + i(m+1)f_m = 0.
\end{equation}
Comparing with Eq.~(\ref{eq:ts}) allows us to relate the decay
exponent $a$ of the velocity amplitude to the order $m$ of the Moore-Saffman function,
\begin{equation}
a=\frac{m+1}{3}.
\label{eq:a}
\end{equation}
Any localized wave motion can be represented as a sum of
Moore--Saffman functions of different orders $m$, each leading to
a wave component characterized by a distinct decay exponent
(\ref{eq:a}). This decay exponent agrees with the derivation of
Peat\cite{Peat1978} for $m=1$, and with the case $m=0$ discussed
by Rieutord {\it et al.}\cite{Rieutord2001} in the problem of
detached layers from critical latitudes in a rotating spherical
shell. Equation~(\ref{eq:a}) is also consistent with the
derivation of Voisin\cite{Voisin2003} for internal waves in a
stratified fluid (the derivation given in this reference is for
the conical wavepacket emitted by a point source, yielding a modified
decay exponent $a_{axi} = a+2/3$).

For a wave beam of order $m$, we can write explicitly the velocity
component along the wave beam $u_x=\Re(U)$ as
\begin{equation}
u_x^{(m)}(x,\eta,t) = |\tilde{U}_0| \left( \frac{\ell}{x} \right)^{(m+1)/3} (c_m (\eta)\cos (\sigma t + \alpha) + s_m (\eta)\sin (\sigma t + \alpha)),
\label{eq:ux}
\end{equation}
and the vorticity component $\omega_y=\Im(W)$ as
\begin{equation}
\omega_y^{(m)}(x,\eta,t) = \frac{|\tilde{U}_0|}{\ell} \left( \frac{\ell}{x} \right)^{(m+2)/3} (-s_{m+1}(\eta) \cos (\sigma t + \alpha) + c_{m+1}(\eta) \sin (\sigma t + \alpha)),
\label{eq:wy}
\end{equation}
where $\tilde{U}_0 = |\tilde{U}_0|e^{-i\alpha}$. The argument $\alpha$ accounts
for a possible phase shift, through added mass effects, between
the wave beam oscillation and the source oscillation. These
quantities (\ref{eq:ux}) and (\ref{eq:wy}) are of interest for
the experimental measurements based on two-component particle
imaging velocimetry in the $(X,Z)$ plane described in Sec.~III.

\subsection{Conservation laws}\label{sec:MOandMS}

We demonstrate now that the order $m$ of the Moore--Saffman
function describing a wave beam  in the far field coincides with
the multipole order $n$ of the source from which it is emitted.
We define in the following a source of order $n$ such that
the moments of order $s$ of the rate of expansion $\mu(X,Z,t) =
\nabla\cdot\mathbf{u}$ are zero for $s<n$ and finite for $s=n$.
We note first that the moment of order $s$ of the
Moore-Saffman function $f_m(\eta)$ (\ref{eq:fm}) satisfies the property
\begin{equation}
  \int_{-\infty}^\infty \eta^s f_m(\eta)\,d\eta =
  \begin{cases}
    0 & \text{($s < m$)}, \\
    i^s\pi s! & \text{($s = m$)}, \\
    \infty & \text{($s > m$)},
  \end{cases}
  \label{eq-tsintprop}
\end{equation}
so that only the $m$-th moment of the velocity profile of order $m$
(\ref{eq:ux}) is finite and non-zero.  What is needed in addition is to find
a conserved quantity involving that moment, and to identify $m$
to the multipole order
of the source from which the wave beam is emitted.

Consider first a line monopole ($n = 0$) along the $Y$-axis,
releasing fluid at the flowrate $q(t)$ per unit length. The
corresponding rate of expansion $\mu(X,Z,t)$ has $q(t)$ as its zeroth moment,
\begin{equation}
  q(t) = \int\mu(X,Z,t)\,dXdZ,
	\label{eq:monoqt}
\end{equation}
and is of the form
\begin{equation}
  \mu(X,Z,t) =
  q(t)
  \delta(X)\delta(Z),
\end{equation}
with $\delta$ the Dirac delta function.  For a domain $D$ of
boundary $C$ in the $(X,Z)$-plane, we have, by the divergence
theorem,
\begin{equation}
  \oint_C \mathbf{u}\cdot\mathbf{n}\,dl
  = \int_D \mu\,dXdZ,
\end{equation}
where $\mathbf{n}$ is the outward normal and $dl$ a positively
oriented contour element. We specialize to the first quadrant and
consider the domain represented in Fig.~\ref{fig:contour}; its
boundary starts from the origin along a segment $C_X$ of the
$X$-axis, continues with a segment $C_z$ perpendicular to the wave
beam at a large distance $x$, and goes back to the origin along a
segment $C_Z$ of the $Z$-axis. The contributions of $C_X$ and
$C_Z$ to the contour integral vanish, since the velocity is
negligible outside the beam. The surface integral is one fourth of
the integral over the whole plane, owing to the parity of the
delta function. We eventually obtain
\begin{equation}
  \int_{-\infty}^\infty u_x\,dz =
  \tfrac{1}{4}q(t),
  \label{eq-monocons}
\end{equation}
a conservation equation of the type used by Moore and
Saffman\cite{Moore1969} and Rieutord \emph{et
al.},\cite{Rieutord2001} involving the zeroth moment of the
longitudinal velocity.

Consider next a line dipole ($n = 1$), defined such that the
flowrate (zeroth moment of the rate of expansion $\mu$) is zero, but
the first moment
\begin{equation}
  \mathbf{p}(t) =
  \int \mathbf{x} \,\mu(X,Z,t)\,dXdZ,
\end{equation}
is finite: it is related to the momentum per unit length imparted
to the fluid. The rate of expansion $\mu$ therefore writes
\begin{equation}
  \mu(X,Z,t) =
  -\mathbf{p}(t)\cdot\nabla(\delta(X)\delta(Z)),
\end{equation}
as discussed for example by Pierce.\cite{Pierce1989}  We
write, with $x_i$ an arbitrary coordinate in the plane $(X,Z)$
and $u_i$ the associated
velocity component,
\begin{equation}
  \nabla\cdot(x_i\mathbf{u}) =
  x_i(\nabla\cdot\mathbf{u}) +
  u_i.
\end{equation}
Integration over an arbitrary  domain $D$ yields
\begin{equation}
  \oint_C x_i\mathbf{u}\cdot\mathbf{n}\,dl =
  \int_D x_i\mu\,dXdZ+
  \int_D u_i\,dXdZ,
\end{equation}
which becomes, after application to the domain $D$ of
Fig.~\ref{fig:contour},
\begin{equation}
  \int_{-\infty}^\infty x_i u_x\,dz =
  \tfrac{1}{4}p_i+\int_0^\infty\!\!\!\int_0^\infty u_i\,dXdZ.
  \label{eq-dipintcons}
\end{equation}
We finally choose $x_i = z$, the cross-beam coordinate, such that $u_i =
u_z$ is negligible everywhere in the quadrant. The last integral
vanishes in Eq.~(\ref{eq-dipintcons}) and we obtain
\begin{equation}
  \int_{-\infty}^\infty zu_x\,dz =
  \tfrac{1}{4}p_z(t),
  \label{eq-dipcons}
\end{equation}
a conservation equation involving the first moment of the
longitudinal velocity. This equation is equivalent to those
used by Thomas and Stevenson\cite{Thomas1972} and
Peat,\cite{Peat1978} involving the zeroth moments of the pressure
and stream function, respectively.
We conclude that the velocity profile in a wave beam emitted by
a line dipole ($n=1$) has a vanishing moment of order $s=0$ (\ref{eq-monocons}) and a finite moment of order $s=1$ (\ref{eq-dipcons}).

The previous argument can be generalized to 
a line source of arbitrary multipole order: the wave beam emitted from a source
of order $n$ is such that the $s$-th moment of the
longitudinal velocity is zero for $s<n$ and finite for $s=n$.
This finite moment writes (see Appendix~\ref{app:a})
\begin{equation}
  \int_{-\infty}^\infty z^n u_x\,dz =
  \tfrac{1}{4}q_{z\cdots z}(t),
  \label{eq-multcons}
\end{equation}
with $q_{z\cdots z}$ the $n$-th moment of the rate of expansion
$\mu(X,Z,t)$ along the $z$ axis.  Switching to complex notation, $u_x^{(m)} =
\Re(U_0e^{-i\sigma t})$ and $q_{z\cdots z} = \Re(Q_{z\cdots
z}e^{-i\sigma t})$, and using Eq.~(\ref{eq:u0a}) yield
\begin{equation}
  \tilde{U}_0 \ell^{1+n} \left(\frac{\ell}{x} \right)^{(m-n)/3}
  \int_{-\infty}^\infty \eta^n f_m(\eta)\,d\eta =
  \tfrac{1}{4}Q_{z\cdots z}.
\end{equation}
Using the property (\ref{eq-tsintprop}) of $f_m(\eta)$ yields immediately $m=n$ and
\begin{equation}
  \tilde{U}_0 = \frac{(-i)^n}{n!}\frac{Q_{z\cdots z}}{4\pi\ell^{n+1}},
  \label{eq-order-exponent}
\end{equation}
confirming that the order $m$ of the Moore--Saffman wave beam is
equal to the multipole order $n$ of the source from which it is
emitted.

We can conclude that a beam emitted by a monopole source
is essentially  an oscillating jet of non-zero instantaneous
flowrate. On the other hand, a wave beam emitted by a multipolar
source of order $m>0$ contains a set of oscillating shear layers with
zero instantaneous flowrate, the number of
layers in the beam increasing as $m^{1/3}$ for large $m$. The
stronger shear stress induced by the larger number of layers
naturally results in a steeper decay of the wave amplitude.

For a source of arbitrary shape, characterized by an arbitrary
multipole expansion, the viscous decay of the wave beams in the
far-field is dominated by the smallest decay exponent $a$, i.e.\
by the term of lowest order in the expansion. In practice we can
focus on the first two orders: monopole ($m=0$) for any source of
finite flowrate, for which $u_x \sim x^{-1/3}$ and $\omega_y \sim
x^{-2/3}$, and dipole ($m=1$) for a source of zero flowrate, for
which $u_x \sim x^{-2/3}$ and $\omega_y \sim x^{-1}$.

\section{Experimental setup and data analysis}\label{sec:setup}

\begin{figure}[t]
\centerline{\includegraphics[width=\columnwidth]{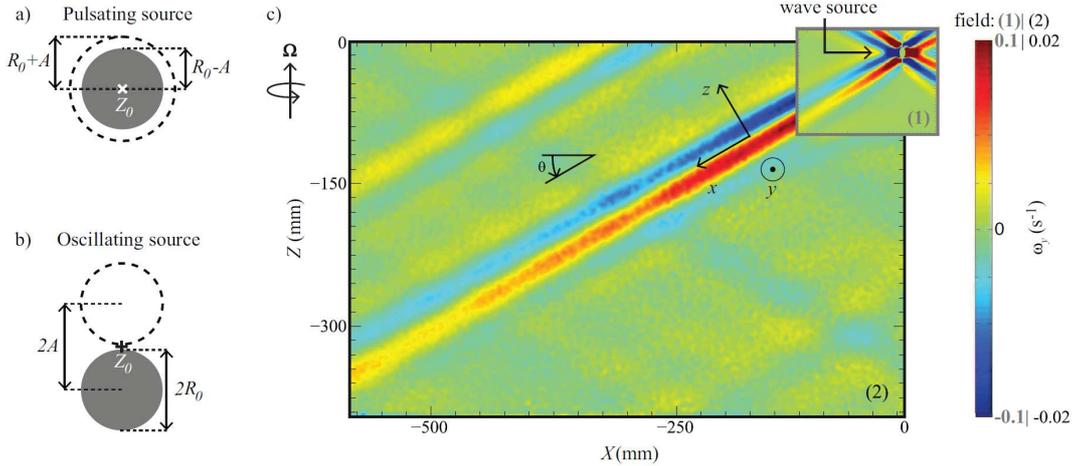}}
\caption{(Color online) Schematic cross-section of the pulsating
(a) and oscillating (b) sources (the two extreme states of the
oscillation are shown). (c) Vorticity component $\omega_y$ showing
the wave beam emitted by the oscillating source, measured by the
two cameras with different resolutions
($\Omega=0.68$~rad~s$^{-1}$, propagation angle
$\theta=\cos^{-1}(\sigma/2\Omega)\simeq 30$\degre). Different
color scales are used for the two fields for better visibility.}
\label{fig:setup}
\end{figure}

We have set up an experiment to characterize the influence of
the multipole order of the source on the structure and
decay of the inertial wave beam.
Measurements are performed in a tank of horizontal size
$L_X \times L_Y=150\times 80$~cm$^2$, filled with 50~cm of water,
and mounted on a 2~m diameter platform
rotating around the vertical axis $Z$. Two wavemakers are
considered, referred to as {\it pulsating source} and {\it
oscillating source} [see Fig.~\ref{fig:setup}(a,b)]. These
wavemakers aim to produce effective monopole and dipole
sources (properties summarized in Tab.~\ref{table:exp_res}):

\begin{itemize}

\item The {\it pulsating source} consists in a water-filled
horizontal rubber tube, 60~cm long, whose volume varies as
$\sin(\sigma t)$. The instantaneous radius varies approximately as
$R(t) \simeq R_0 + A\sin(\sigma t)$, with mean radius $R_0=8.9$~mm
and amplitude $A=0.7$~mm (the oscillation is harmonic to within
$A/R_0 \simeq 8$\%). This source imposes an oscillating flow rate
per unit length $q(t) \simeq 2\pi R_0 A \sigma \cos (\sigma t)$.

\item The {\it oscillating source} consists in a horizontal
cylinder, 60~cm long, $R_0=3.0$~mm in radius, whose vertical
position $Z(t)=Z_0+A\sin(\sigma t)$ oscillates at frequency
$\sigma$ and amplitude $A=3.2$~mm (Fig.~\ref{fig:setup}(b)). This
cylinder is a source of zero net flowrate which can be modeled, at
large distances, as an oscillating dipole characterized by a
dipole moment $\mathbf{p}(t) = \Re(\mathbf{P}e^{-i\sigma t})$,
with $\mathbf{P} = [1+C(\sigma)] \pi R_0^2 A \sigma \mathbf{e}_Z$
and $C(\sigma)$ an added mass coefficient. Without background
rotation, this coefficient is a real constant $C = 1$. With
rotation, the coefficient becomes complex and frequency-dependent
owing to wave generation. The experimental measurement and
theoretical determination of added mass coefficients have been
considered by Ermanyuk and Gavrilov\cite{Ermanyuk2002a} and
Ermanyuk,\cite{Ermanyuk2002b} among others, for internal waves.

\end{itemize}

\begin{table}[b]
\centering
\begin{tabular}{p{2.8cm}|p{1.5cm}|p{1.5cm}|p{1.5cm}|p{1.5cm}|p{2cm}|p{2cm}}
\centering Source & \centering $R_0$ & \centering $A$ & \centering $A\sigma$ & \centering $Re$ & \multicolumn{2}{p{4cm}}{\centering Vorticity decay exponent} \\
&  \centering [mm] & \centering [mm] & \centering [mm~s$^{-1}$] &  & \centering Theory & \multicolumn{1}{p{2cm}}{\centering Experiment}\\
\hline \hline
\centering Pulsating ($m=0$) & \centering 8.9 & \centering 0.7 & \centering 0.82 & \centering 15 & \centering $-2/3$ & \multicolumn{1}{p{2cm}}{\centering $-0.64\pm0.09$}\\
\hline \centering Oscillating ($m=1$) & \centering 3.0 &
\centering 3.2 & \centering 3.77 & \centering 23 & \centering $-1$
& \centering $-0.99\pm0.05$
\end{tabular}
\caption{Properties of the pulsating  and oscillating sources:
mean radius $R_0$, oscillation amplitude $A$, velocity amplitude
$A\sigma$, Reynolds number $Re=2 R_0 A\sigma/\nu$ (with
$\nu=10^{-6}$~m$^2$~s$^{-1}$ the kinematic viscosity of water).
The vorticity decay exponents correspond to the average ($\pm$
standard deviation) over the four experiments at different
$\sigma/2\Omega$ [see
Fig.~\ref{fig:decay}(b)].}\label{table:exp_res}
\end{table}

The wavemaker (either pulsating or oscillating) is immersed
horizontally 10~cm below the surface.
It is located along the $Y$ axis, at a distance $\Delta X=30$~cm from the
side wall of the tank. The wavemaker frequency is
kept constant, $\sigma=1.18$~rad~s$^{-1}$, so that the
wavemaker velocity $A\sigma$ is constant. The rotation rate of the
platform $\Omega$ is varied in the range 0.68 to 1.68 rad~s$^{-1}$
(6.5 to 16~rpm), resulting in a beam angle $\theta =
\cos^{-1}(\sigma/2\Omega)$ varying in the range $30-70^\mathrm{o}$.

The Reynolds number of the
flow in the vicinity of the wavemaker, defined as $Re=2 R_0
A\sigma/\nu$, is $Re=15$ for the pulsating source and $Re=23$ for
the oscillating source. Despite the relatively large amplitude
ratio (up to $A/R_0 = 1.06$ for the oscillating cylinder), these moderate
Reynolds numbers indicate that nonlinearities
(saturation and generation of higher harmonics) can be neglected;
see in particular the discussion by Voisin {\it et al.}\cite{Voisin2011}
for internal waves, pointing the importance of
$(A/R_0)\mathit{Re}$ for saturation. 
The viscous
scale $\ell$ (\ref{eq:ell}) varies in the range $0.6$--$1.2$~mm.
For the pulsating cylinder, the ratio $R_0 / \ell \simeq 7-14$
indicates that the far-field properties of
the wave beams are expected at a significant distance $x/\ell$
(the radius of the oscillating
cylinder can be made arbitrarily small, but the radius of the
pulsating cylinder is limited by the design of the rubber tube).

The two components of the velocity fields $(u_X,u_Z)$ are measured
in the vertical plane $Y=L_Y/2$ normal to the source axis using a
particle image velocimetry (PIV) system mounted in the rotating
frame. Among the four wave beams emitted by the sources, we focus
on the one propagating over the longest distance, in the
bottom-left direction [see Fig.~\ref{fig:setup}(c)]. Images of
particles are acquired with two $2360 \times 1776$~pixels cameras
operating simultaneously with different fields of view. Each PIV
acquisition consists in 3\,000 image pairs (one image per camera)
recorded at 3~Hz, which represents 16 fields per source period.
Cross-correlation between successive images produces velocity
fields sampled on a grid of $295\times 222$ vectors, with a
spatial resolution of 0.58~mm for the closer view and 2.04~mm for
the larger view. The combination of the PIV data from the two
cameras allows us to resolve accurately the spatial scales of the
wave field for distances from the source $x$ between 10~mm and
1~m.

Two post-processing steps are applied to the PIV fields. First,
the velocity time series are phase-averaged at the forcing
frequency $\sigma$ in order to filter out contributions from
unwanted residual flows. These residual flows originate from
thermal convection effects (velocities of the order of
1~mm~s$^{-1}$), at very small frequencies, and motions due to the
precession of the rotating platform induced by the Earth
rotation\cite{Boisson2012a} ($\simeq 0.5$~mm~s$^{-1}$), at
frequency $\Omega$. Second, we apply a spatial Fourier filter to
remove flow structures associated to wave vectors $\mathbf{k}$
such that $k_X\,k_Z>0$. This procedure is useful to remove
secondary wave beams reflecting on the tank walls, which intersect
the primary beam characterized by $k_X\,k_Z<0$ and induce spatial
oscillations of the wave envelope [see Fig.~\ref{fig:setup}(c)].

Finally, we remap the velocity fields in the tilted frame $(x,z)$
of the beam, with $x$ the distance from the source, making the
angle $\theta=\cos^{-1}(\sigma/2\Omega)$ to the horizontal, and
compute the out-of-plane vorticity component $\omega_y$.
A standard second-order finite difference
scheme is used to compute $\omega_y$, which is comfortably
resolved by our twin PIV measurements ensuring at least 40 grid
points per wavelength at all distance $x$ from the source. The
vorticity and velocity envelopes of the wave field are finally
computed as $\omega_0(x,z)=\langle 2\omega_y^2\rangle^{1/2}$ and
$u_0(x,z)=\langle 2u_x^2\rangle^{1/2}$, with $\langle \rangle$ a
temporal average.

\section{Results}\label{sec:results}

\begin{figure}
\centerline{\includegraphics[width=\columnwidth]{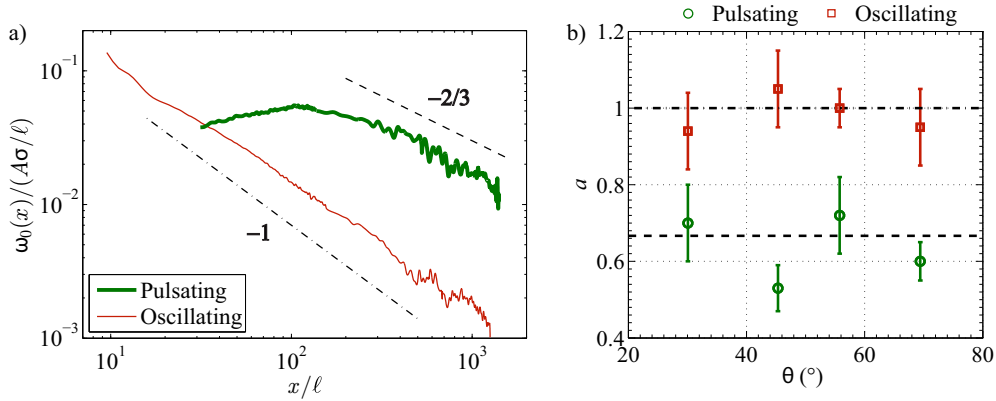}}
\caption{(Color online) (a) Vorticity envelope $\omega_0(x,z=0)$
along the direction of the wave beam for the pulsating (thick
line) and oscillating (thin line) sources
($\Omega=1.05$~rad~s$^{-1}$, $\theta\simeq56$\degre, $\ell =
0.76$~mm). The lines show power laws of exponents $-2/3$ and $-1$
expected theoretically for a monopole and a dipole source
respectively. (b) Vorticity decay exponent as a function of the
propagation angle $\theta=\cos^{-1}(\sigma/2\Omega)$:
($\medcirc$), pulsating source; ($\Square$), oscillating
source.}\label{fig:decay}
\end{figure}

We first compare the spatial decay of the vorticity envelope of
the wave beam for the pulsating and oscillating sources. Vorticity
is used here instead of velocity to compare against the theory
(\ref{eq:wy}) because it is less sensitive to residual large scale
flows. The centerline vorticity $\omega_0(x,z=0)$ is plotted as a
function of the distance $x$ from the source in
Fig.~\ref{fig:decay}(a) for $\Omega=1.05$~rad~s$^{-1}$
($\theta\simeq56$\degre). The distance is normalized by the
viscous length $\ell$, and vorticity $\omega_0$ is normalized by
the source velocity $A \sigma$ and $\ell$. For both sources a
power law decay of the vorticity is observed far from the source,
but with a different exponent: $\omega_0 \sim x^{-1.00 \pm 0.05}$
for the oscillating source, starting close to the source ($x/\ell
> 10$) and extending over nearly two decades; $\omega_0 \sim
x^{-0.72 \pm0.10}$ for the pulsating source, starting much
further from the source ($x/\ell > 200$) and hence visible over a
limited range of $x$. These exponents are in good agreement with
the predictions for a monopole and a dipole source, $x^{-2/3}$ and
$x^{-1}$ respectively [see Eq.~(\ref{eq:wy})]. Exponents measured
at other values of $\sigma/2\Omega=\cos\theta$, reported in
Fig.~\ref{fig:decay}(b), are consistent with these numbers and do
not show any trend with $\theta$.

\begin{figure}
\centerline{\includegraphics[width=.7\columnwidth]{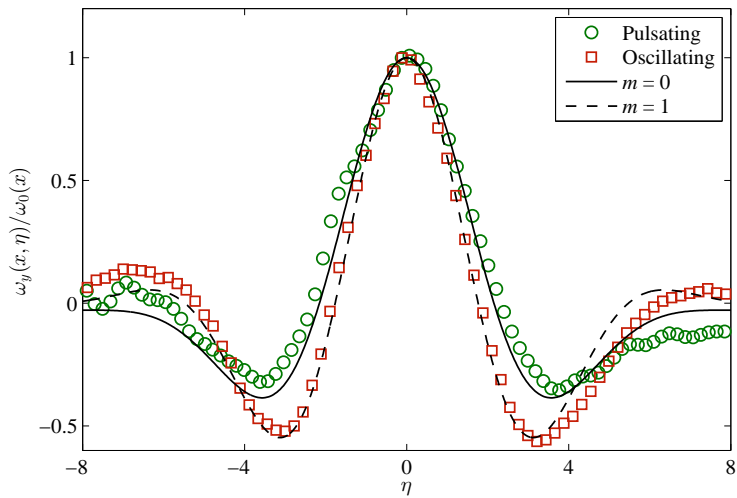}}
\caption{(Color online) Normalized transverse vorticity profiles
$\omega_y(x,\eta,t)/\omega_0(x,0)$ as a function of the reduced
transverse coordinate $\eta = z/(x^{1/3} \ell^{2/3})$, compared
to the theoretical vorticity profiles $c_{m+1}(\eta)/c_{m+1}(0)$.
$\medcirc$, pulsating source ($m=0$); $\Square$, oscillating source ($m=1$).
The profiles are measured at a distance $x/\ell=330$ from the source, in the region where both
beams show a power law decay, and are shown for the time at which the vorticity is
maximum at the center of the beam.}
\label{fig:prof_inst}
\end{figure}

Another distinctive property of a wave beam emitted by a monopole
or a dipole disturbance is the transverse vorticity profile, given
by the function $f_1(\eta)$ or $f_2(\eta)$ respectively.
Figure~\ref{fig:prof_inst} shows the normalized vorticity profile
$\omega_y(x,z,t)/\omega_0(x,z=0)$ for both sources as a function
of the reduced transverse coordinate $\eta = z/(x^{1/3}
\ell^{2/3})$, at the time at which $\omega_y$ is maximum at the
centerline. The normalized profiles, shown here for a distance
$x/\ell=330$, are independent of $x$ provided that $x$ is large
enough, i.e.\ in the region showing a well-defined power-law decay
($x/\ell>200$ for the pulsating source and $x/\ell>10$ for the
oscillating source). We compare these profiles against the
Moore--Saffman functions taken at the same phase,
$\omega_y(x,\eta,t)/\omega_0(x,\eta=0)= c_{m+1}(\eta)/c_{m+1}(0)$, for $m=0$
and $m=1$. The agreement is excellent for both sources, for
$|\eta|$ up to $5$, clearly confirming that the order of the
Moore--Saffman function that best describes the wave envelope is
governed by the multipole order of the source. At larger distance
from the beam centerline ($|\eta|>5$), the discrepancy between the
theoretical and experimental profiles probably originates from
residual fluid motions associated to reflected wave beams that
cannot be eliminated by the Fourier filtering procedure.

The flowrate across the wave beam provides another confirmation of
the match between the wave beam order and the source multipole
order. We compute the flowrate $q(t)$ per unit length by
integration over $\eta$ of the instantaneous velocity
$u_x(x,\eta)$, using a truncation at $|\eta|=6$ to reduce
disturbance from fluid motions out of the primary beam of
interest. The theoretical flowrate amplitude for each wave beam
is $q_{th} = \pi \ell^{2/3} x^{1/3} u_0(x,\eta=0)$ for the monopole source and zero
for the dipole source. We find a normalized flowrate $q/q_{th}
\simeq 1.0 \pm 0.05$ for  the
pulsating cylinder, and $0.07 \pm 0.05$ for the oscillating cylinder
(the non-zero value in the latter case originates from the
truncation of the integral). This normalized flowrate
does not depend significantly on the distance from the source in
the far-field for both disturbances (to within $\pm 0.05$).

\begin{figure}
\centerline{\includegraphics[width=.7\columnwidth]{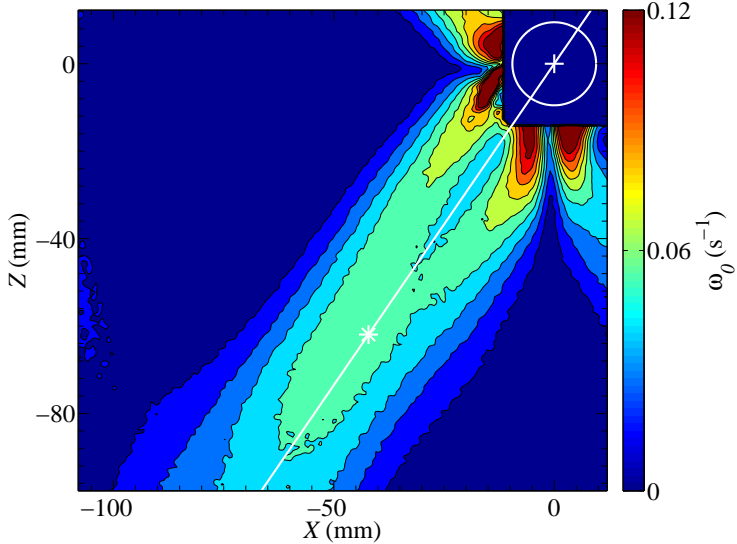}}
\caption{(Color online) Contour plot of the vorticity envelope
$\omega_0$ from the close-view camera for the pulsating source at
$\theta = 56^\mathrm{o}$, showing the transition from a bimodal
to a unimodal wave beam. The white circle shows the source and
the white line the far-field wave beam axis. The star marker
indicates the location $x_c/\ell=110$ at which $\omega_0(x)$ is
maximum in Fig.~\ref{fig:decay}(a).} \label{fig:envelope}
\end{figure}

We finally turn to the description of the distance $x$ beyond
which the scaling law holds for the far-field decay of the wave
amplitude. Figure~\ref{fig:decay}(a) shows that the vorticity
amplitude decreases at all $x$ for the oscillating source, with a
well-defined power law beyond $x/\ell \simeq 10$, while it is
non-monotonic for the monopole source with a maximum at $x_c /
\ell \simeq 110 \pm 10$. This non-monotonic profile originates
from the transition from a bimodal beam close to the source to a
unimodal beam far from the source, as illustrated by the close-up
view in Fig.~\ref{fig:envelope}. Because of the large extent of
the pulsating cylinder compared to the viscous length, the wave
field close to the source corresponds to two separate beams
propagating in the same direction
$\theta=\cos^{-1}(\sigma/2\Omega)$, with a shift of
$z=\pm R_0=\pm 8.9$~mm from the axis of the far-field beam. Close
to the source, the transverse profile is therefore bimodal, with a
local minimum at $z=0$, while it becomes unimodal for large
distances with a maximum at $z=0$, with a transition occurring
around $x_c$.

\begin{sloppypar}
This bimodal-to-unimodal transition, a classical feature of waves
emitted from a source of large extent, has been mainly described
for internal waves excited by oscillating
disturbances.\cite{Makarov1990,Hurley_Keady_97_2,Flynn2003,Zhang2007,Gostiaux_thesis,Voisin2011}
Figure~\ref{fig:sketch}(a,b) provides a simple sketch of this
transition for a pulsating and an oscillating cylinder of large
extent. The oscillating boundary layer over the cylinder detaches
at the four critical lines, where the local slope equals the wave
beam angle $\theta$, forming eight wave beams of non-zero
instantaneous flowrate.
The far-field beam in a given direction
results from the merging of two parent beams propagating along
the same direction, that are in phase for the pulsating cylinder
and out-of-phase for the oscillating cylinder. The combination of
two in-phase parent beams of order $m$ separated by a
normalized transverse distance $\eta_0$ is given by
$f_m(\eta+\eta_0/2) + f_m(\eta-\eta_0/2)$, which far from the source
($\eta_0 \ll 1$) simply gives a beam of order $m$. On the other
hand, the combination of two out-of-phase parent beams of
order $m$ is given by $f_m(\eta+\eta_0/2) - f_m(\eta-\eta_0/2)
\simeq \eta_0 f'_m(\eta) = i \eta_0 f_{m+1}(\eta)$, yielding a
beam of order $m+1$: this is consistent with the observation of a
far-field wave beam of zero flowrate ($m=1$) emitted from the
two nonzero flowrate sources ($m=0$) at the critical lines for the
oscillating cylinder.
\end{sloppypar}

Determining the merging distance $x_c$ requires the full
resolution of the flow close to the wavemaker. This computation is
given in Hurley and Keady\cite{Hurley_Keady_97_2} and
Voisin\cite{Voisin2011} for oscillating cylinders and spheres in a
stratified fluid. Qualitatively, we can estimate $x_c$ from the
spreading law $\delta(x) \sim
\ell^{2/3} x^{1/3}$ of each parent beam, yielding a merging distance $x_c$ such
that $\delta(x_c) \simeq 2R_0$ given by $x_c / \ell \sim
(R_0/\ell)^3$. The radius ratio between the pulsating and
oscillating cylinders ($R_0 = 8.9$ and $3.0$~mm, respectively)
indicates that $x_c/\ell$ is expected much larger for the pulsating
cylinder. For the oscillating cylinder the predicted merging
distance is of order of the cylinder radius, so that the bimodal
wave beam cannot be observed, which is consistent
with the monotonic decay of $\omega_0(x)$.

\section{Conclusion}\label{sec:concl}

In this paper we analyzed the decay of a two-dimensional inertial wave beam and
showed that it is set by the multipole order $n$ of the source from which it is emitted.
Experimental measurements are reported for sources of the first two orders:
monopole ($n=0$) and dipole ($n=1$).
We find that the structure of the wave beam is well represented by a
Moore-Saffman\cite{Moore1969} (or Thomas-Stevenson\cite{Thomas1972})
function of order equal to the
multipole order of the source.
The wave envelope decays as a power law of the distance from the
source, with an exponent governed by the
order of the source ($x^{-(n+1)/3}$ for the velocity and $x^{-(n+2)/3}$ for
the vorticity). These properties, demonstrated here
for inertial waves in rotating fluids, should also hold
for internal waves in stratified fluids.

The steeper decay of the wave amplitude as the multipole order
is increased indicates that the far-field structure of a wave beam 
is dominated by the lowest order of the source, i.e. by its
first nonzero moment (flowrate, momentum or higher order moment).
In most practical situations, the nature of the source can be
discriminated by its instantaneous flowrate, either nonzero
for a monopole source or zero for a dipole or higher order source.
This means that a wave beam emitted from any source of nonzero instantaneous
flowrate must be dominated in the far field by its monopole component.
Such monopole sources are relevant to most natural flows (e.g. conical wave beams
in the fluid core of planets, internal tide generation over
ocean topography): the detachment of the oscillating boundary
layers at the critical lines produces oscillating jets in the bulk,
corresponding to a weak spatial decay.
On the other hand,  oscillating disturbances immersed in the fluid (a typical configuration
of most laboratory experiments in rotating or stratified fluids)
produce in the far field beams of zero
instantaneous flowrate, resulting in a stronger spatial decay.

\section*{Acknowledgements}
We acknowledge A. Aubertin, L. Auffray, A. Campagne and R. Pidoux for
experimental help, and M. Rieutord for fruitfull discussions.
This work is supported by the ANR Grant
No.~2011-BS04-006-01 ``ONLITUR''. FM acknowledges the Institut
Universitaire de France for its support, and BV the Labex OSUG@2020
(Investissements d'avenir - ANR10 LABX56).

\appendix

\section{Conservation laws for arbitrary multipole order}
\label{app:a}

Equations (\ref{eq-monocons}) and (\ref{eq-dipcons}) relate the zeroth and first moments of the longitudinal velocity to the flowrate and momentum of monopole ($n=0$) and dipole ($n=1$) sources. In this appendix we generalize these relations to sources of arbitrary multipole order $n$. We introduce arbitrary orthogonal coordinates $(x_1, x_2)$ in the plane $(X,Z)$, with $(u_1, u_2)$ the associated velocity components. The rate of expansion $\mu(X,Z,t)$ of a source of multipole order $n$ has zero moments of order $0$ to $n-1$ and finite $n$-th moments given by
\begin{equation}
  q_{i_1\cdots i_n}(t) =
  \int x_{i_1}\cdots x_{i_n}\mu(X,Z,t)\,dXdZ.
\end{equation}
These $n$-th moments are tensors of rank $n$ composed
of $n+1$ independent scalars.
The rate of expansion is of the form
\begin{equation}
  \mu(X,Z,t) =
  \frac{(-1)^n}{n!}
  \sum_{i_1 = 1}^2\cdots\sum_{i_n = 1}^2
  q_{i_1\cdots i_n}(t)
  \frac{\partial^n}{\partial x_{i_1}\cdots\partial x_{i_n}}
  \delta(X)\delta(Z).
\end{equation}
Using the identity
\begin{equation}
  \nabla\cdot(x_{i_1}\cdots x_{i_n}\mathbf{u}) =
  x_{i_1}\cdots x_{i_n}(\nabla\cdot\mathbf{u}) +
  u_{i_1}x_{i_2}\cdots x_{i_n} + \cdots + x_{i_1}\cdots x_{i_{n-1}}u_{i_n},
\end{equation}
and integrating over the domain $D$ of Fig.~\ref{fig:contour} yield
\begin{multline}
  \int_{-\infty}^\infty x_{i_1}\cdots x_{i_n}u_x\,dz =
    \tfrac{1}{4}q_{i_1\cdots i_n}
    \\
  \mbox{}+
  \int_0^\infty\!\!\!\int_0^\infty
    (u_{i_1}x_{i_2}\cdots x_{i_n} + \cdots + x_{i_1}\cdots x_{i_{n-1}}u_{i_n})\,dXdZ.
		\label{eq:lastint}
\end{multline}
We choose $(x_1,x_2) = (x,z)$ and $i_1 = \cdots = i_n = 2$, so that  $x_{i_1} = \cdots = x_{i_n} = z$. The corresponding transverse velocity $u_{i_1} = \cdots = u_{i_n} = u_z$ is negligible, so the last integral vanishes in Eq.~(\ref{eq:lastint}). We finally obtain Eq.~(\ref{eq-multcons}), which generalizes Eqs.~(\ref{eq-monocons}) and (\ref{eq-dipcons}) to arbitrary multipole order $n$.


\begin{thebibliography}{99}
\bibitem{footnote}
The decay exponent $a=1/3$ discussed by Cortet {\it et al.}~\cite{Cortet2010} corresponds to the case of a monopole source, although the wavemaker used in that reference (oscillating cylinder) should behave as a dipole source, for which a steeper decay is expected ($a=2/3$).
Because of the restricted range of distance from the source, the difference between the two exponents could not be discriminated by the data.
\end{thebibliography}
\end{document}